\documentclass[twocolumn,aps,prl]{revtex4}
\usepackage{graphicx}

\usepackage{pdfpages}

\begin{document}

\title{Revealing sign-reversal $s^{+-}$-wave pairing by quasiparticle interference in the heavy-fermion superconductor CeCu$_2$Si$_2$}

\author{Shan Zhao}
\affiliation{Department of Physics, Beijing Jiaotong University, Beijing 100044, China}

\author{Bin Liu}
\email[]{liubin@bjtu.edu.cn}
\affiliation{Department of Physics, Beijing Jiaotong University, Beijing 100044, China}

\author{Yi-feng Yang}
\affiliation{Beijing National Laboratory for Condensed Matter Physics,  Institute of Physics,
Chinese Academy of Science, Beijing 100190, China}
\affiliation{School of Physical Sciences, University of Chinese Academy of Sciences, Beijing 100190, China}
\affiliation{Songshan Lake Materials Laboratory, Dongguan, Guangdong 523808, China}

\author{Shiping Feng}
\affiliation{Department of Physics, Beijing Normal University, Beijing 100875, China}

\date{\today}

\begin{abstract}
Recent observations of two nodeless gaps in superconducting CeCu$_2$Si$_2$ have raised intensive debates as to its exact gap structure of either sign-reversal ($s^{+-}$) or sign-preserving ($s^{++}$) pairing. Here we investigate the quasiparticle interference (QPI) using realistic Fermi surface topology for both weak and strong interband impurity scatterings. Our calculations of the QPI and integrated antisymmetrized local density of states reveal qualitative distinctions between $s^{+-}$ and $s^{++}$ pairing states, which include the intragap impurity resonance and a significant energy-dependence difference between two gap energies. Our predictions provide a guide for phase-sensitive QPI measurements to uncover decisively the true pairing symmetry in the heavy-fermion superconductor CeCu$_2$Si$_2$.

\end{abstract}

%\pacs{74.70.Tx, 74.20.Pq, 74.55.+v, 74.62.En}
\maketitle

The past decades have witnessed continuing debates as to the superconducting (SC) pairing symmetry of heavy-fermion superconductors \cite{Stewart1984,Vojta2007,Si2010,Kirchner2020}. Despite tremendous experimental efforts, unambiguous detection of their SC gap structures has proved difficult, mostly because of their extremely small energy scales ($\sim1$ meV) and low transition temperatures ($T_c\sim1$ K). Direct measurement and exact mapping of the electronic band structures using angle-resolved photoemission spectroscopy (ARPES) are severely limited by energy resolution. Proposals of the SC gap structures and pairing symmetry are thus largely derived from indirect evidence such as specific heat, nuclear magnetic resonance (NMR), neutron scattering and so on, which inevitability causes discrepancies and controversies. CeCu$_2$Si$_2$ is one of the most noticeable examples.

As the first unconventional superconductor, the superconductivity in CeCu$_2$Si$_2$ was initially discovered in 1979 \cite{Steglich1979} and has long been believed to be of $d$-wave Cooper pairing mediated by antiferromagnetic spin fluctuations at ambient pressure \cite{Stockert2011,Eremin2008,Fujiwara2008,Ishida1999,Vieyra2011}. This physical picture, however, has been questioned since 2014 by a number of refined experiments including angle-resolved specific heat \cite{Kittaka2016}, London penetration depth \cite{Pang2016,Takenaka2017}, and thermal conductivity \cite{Yamashita2017}, measured down to rather low temperatures on high quality samples. All of these experiments support multiband superconductivity with two nodeless gaps. Both sign-reversal $s^{+-}$-wave pairing and sign-preserving $s^{++}$-wave pairing have been proposed to fit the experimental data. As a consequence, intensive debates arise as to which the true pairing symmetry is.

Similar to iron pnictides, the spin-fluctuation-mediated $s^{+-}$ wave is the most competitive candidate for multiband superconductivity in CeCu$_2$Si$_2$ \cite{Ikeda2015,Yang2018}. The opposite sign structure between electron and hole Fermi surfaces could explain the neutron spin resonance mode below $T_c$ \cite{Stockert2011}, as well as the $T^3$ behavior and the absence of the Hebel-Slichter peak in the spin-lattice relaxation rate as long as the interband coherence factor is considered \cite{Kitaoka1986,Fujiwara2008}. On the other hand, a recent electron irradiation experiment reported robust superconductivity against impurities and supported the $s^{++}$ wave without sign change \cite{Yamashita2017}. The $s^{++}$-wave pairing could be favored if orbital fluctuations are dominant \cite{Nica2017}, but one should be reminded of earlier observations that $T_c$ was significantly suppressed by impurity \cite{Spille1983,Adrian1987,Yuan2004}. Thus, further experimental confirmations are needed and the determination of the gap symmetry and its exact sign structure is of utmost importance as a clue to uncover the SC mechanism in CeCu$_2$Si$_2$.

Unfortunately, direct experimental probes on the gap structure of superconducting CeCu$_2$Si$_2$ are still lacking up to the present. Conventional phase-sensitive measurements developed to identify the $d$-wave pairing in cuprates have proved useless in distinguishing the $s^{+-}$- and $s^{++}$-wave pairing states in iron-based superconductors. In this respect, a promising alternative technique was recently proposed based on phase-sensitive quasiparticle interference (QPI) or Fourier transform scanning tunneling microscopy (FT-STM). Its application in LiOH-intercalated FeSe \cite{Du2017} and NaFe$_{1-x}$Co$_x$As \cite{Cheung2020} has revealed qualitative differences in the integrated antisymmetrized intensities of the local density of states (LDOS) to determine the sign-reversal order parameter \cite{Hirschfeld2015,Hirschfeld2018,Gao2018}.

In this paper, we make theoretical predictions on the QPI with both intraband and interband impurity scatterings in CeCu$_2$Si$_2$. We find that for strong interband impurity scattering, only the $s^{+-}$-wave pairing shows the well-known intra-gap resonance states in both LDOS and integrated antisymmetrized LDOS. The integrated antisymmetrized LDOS always indicates significant energy dependence, but there exist qualitative differences between two gap scales for the $s^{+-}$ and $s^{++}$ pairing symmetries irrespective of impurity scattering strength. These robust features can be used to unambiguously identify the nodeless $s^{+-}$-wave gap and provide a useful guide for the phase-sensitive QPI or FT-STM  to solve the highly debated issue of pairing symmetry in CeCu$_2$Si$_2$.

\begin{figure}[t]
\includegraphics[width=0.48\textwidth]{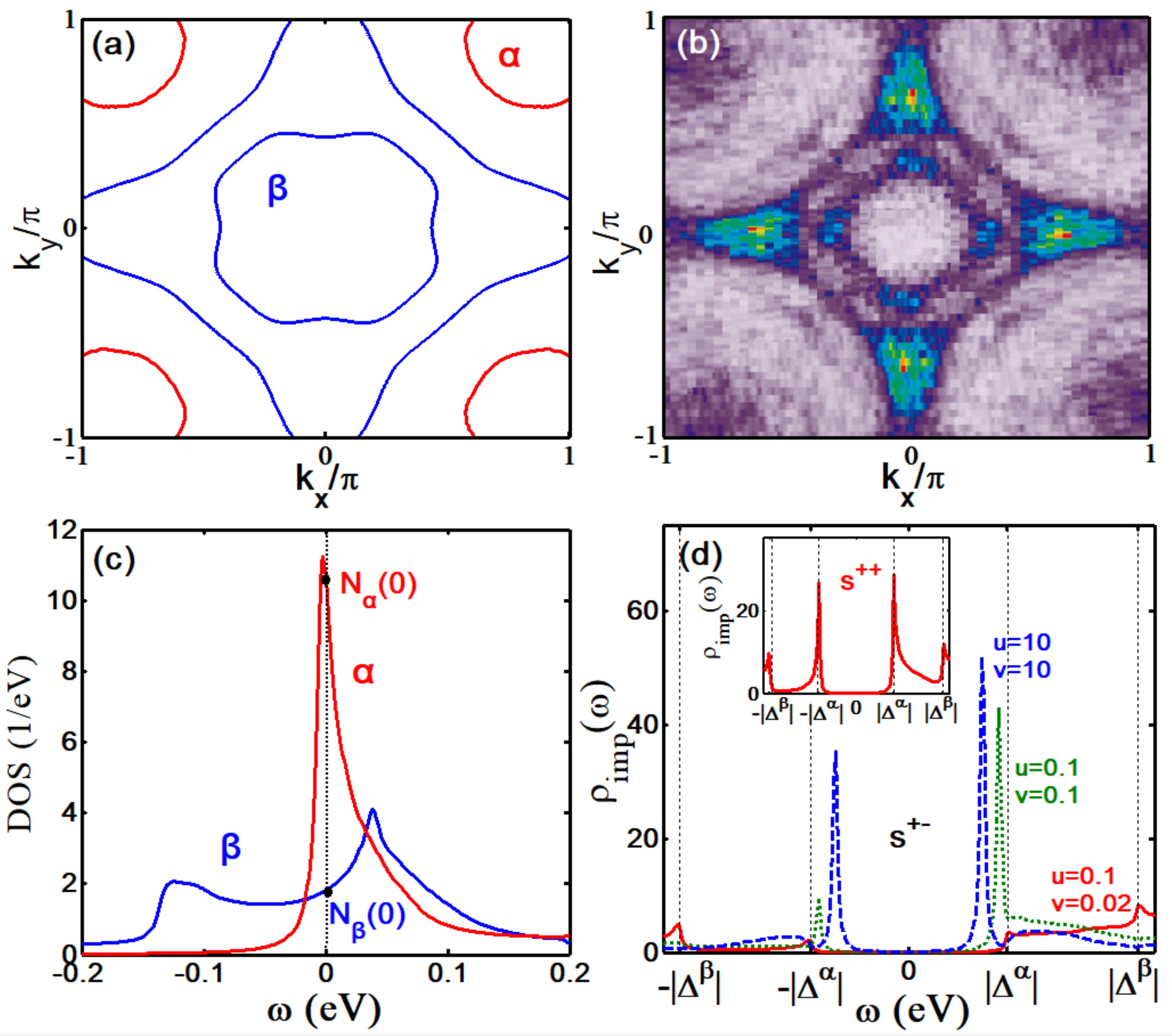}
\caption{(a) Calculated two-dimensional Fermi surfaces, composed of the heavy-electron Fermi surface $\alpha$ around $(\pi,\pi)$ and the hole Fermi surfaces $\beta$ around $(0,0)$ from DFT+$U$ at ambient pressure. (b) Comparison with the ARPES Fermi surfaces extracted from Fig. 4(a) in Ref. [28]. (c) The band-resolved DOS in the normal state for $\alpha$ and $\beta$ bands. The sharp peak in the $\alpha$ band close to the chemical potential reflects its extreme flatness with typical $f$-electron character. (d) The LDOS at a single nonmagnetic impurity for $s^{+-}$-wave pairing with different intraband and interband impurity scattering strength. The inset plots the LDOS for $s^{++}$-wave pairing for comparison.}
\label{fig1}
\end{figure}

The presence of multiple Fermi surfaces is crucial for the multi-gap superconductivity of CeCu$_2$Si$_2$ and has been obtained by us based on density functional theory calculations with Hubbard $U$ correction (DFT+$U$) \cite{Yang2018,Liu2019}. Although the band structures are very complex, it is clearly shown that only two hybridization bands cross the Fermi energy and dominate the essential low-energy physics of CeCu$_2$Si$_2$ \cite{Yang2018,Liu2019}. Therefore we refer to the above effective low-energy hybridization bands as our two-hybridization-band model for investigating the unconventional superconductivity of CeCu$_2$Si$_2$. The resulting three-dimensional Fermi surfaces contain two major parts: the complex inner hole sheets and the corrugated-cylinder heavy-electron sheet barely dispersing along the $k_{z}$ direction. Figure 1(a) plots the typical two-dimensional (2D) mapping, which includes the heavy-electron Fermi pockets around $(\pi,\pi)$, denoted as the $\alpha$ band, and the light-hole Fermi pockets (inner and outer) around $(0,0)$, denoted as the $\beta$ band. The obtained 2D Fermi surfaces fit extremely well with the recent ARPES experiment shown in Fig. 1(b) \cite{Wu2021}. The sharp peak in the density of states (DOS) in Fig. 1(c) close to the chemical potential reflects the extreme flatness of the heavy $\alpha$ band of typical $f$-electron character due to the many-body Kondo effect. At the Fermi energy, the DOS of the $\alpha$ band, $N_\alpha(0)$, is about six times larger than that of the $\beta$ band, $N_\beta(0)$, suggesting $|\Delta^{\beta}/\Delta^{\alpha}|\approx (N_\alpha(0)/N_\beta(0))^{1/2}\approx 2.4$ \cite{Senga2009} when interband pairing interaction is dominant for CeCu$_2$Si$_2$ \cite{Yang2018}. This ratio is in reasonable agreement with experimental estimates \cite{Kittaka2016,Pang2016,Kittaka2014,Enayat2016}. According to self-consistently calculations in Ref. [16], we take $|\Delta^{\beta}/\Delta^{\alpha}|=2.5$ with $|\Delta^{\alpha}|=0.1$ meV in the following calculations \cite{Stockert2011,Yang2018}. These justify the starting point of our calculations based on the effective two-hybridization-band model.

%\begin{figure}[t]
%\includegraphics[width=0.48\textwidth]{fig1.pdf}
%\caption{(a) Calculated two-dimensional Fermi surfaces, composed of the heavy electron Fermi surface $\alpha$ around $(\pi,\pi)$ and the hole Fermi surfaces $\beta$ around $(0,0)$ from DFT+$U$ at ambient pressure. (b) Comparison with the ARPES Fermi surfaces extracted from Fig. 4(a) in Ref. [28]. (c) The band-resolved DOS in the normal state for $\alpha$ and $\beta$ bands. The sharp peak of the $\alpha$-band close to the chemical potential reflects its extreme flatness with typical $f$-electron character. (d) The LDOS at a single nonmagnetic impurity for $s^{+-}$-wave pairing with different intraband and interband impurity scattering strength. The inset plots the LDOS for $s^{++}$-wave pairing for comparison.}
%\label{fig1}
%\end{figure}

For calculations in the SC state, we introduce a four-component Nambu spinor operator so that the bare Green's function can be formulated as
\begin{eqnarray}
\hat{G}^{-1}_{0}({\bf k};i\omega_{n})=i\omega_{n}\hat{1}-\left (\matrix{\varepsilon^{\alpha}_{\bf k} &\Delta^{\alpha}
&0 &0\cr \Delta^{\alpha} &-\varepsilon^{\alpha}_{\bf k}
&0 &0\cr 0 &0 &\varepsilon^{\beta}_{\bf k} &\Delta^{\beta}\cr 0 &0 &\Delta^{\beta} &-\varepsilon^{\beta}_{\bf k}\cr}\right),
\end{eqnarray}
where $\omega_n$ is the fermionic Matsubara frequency and $\varepsilon^{\alpha}_{\bf k}$ and $\varepsilon^{\beta}_{\bf k}$ are two-hybridization-band dispersions from the above DFT+$U$ calculations. Only intraband pairings $\Delta^{\alpha}$ and $\Delta^{\beta}$ are considered here due to the well-separated $\alpha$ and $\beta$ Fermi surfaces in the Brillouin zone. Without loss of generality, we take the nonmagnetic impurity scattering matrix of the following structure:
\begin{eqnarray}
\hat{U}=\left(\matrix{u &0&v &0\cr  0 & -u& 0& -v\cr
v&0 &u &0\cr 0 &-v &0 &-u\cr}\right),
\end{eqnarray}
where $u$ and $v$, in units of eV, are the strength of intraband and interband scattering potentials, respectively. The effect of the impurity scattering can be treated within the $T$-matrix approach, giving \cite{Zhu2006}
\begin{eqnarray}
\hat{T}(i\omega_{n})=\frac{\hat{U}}{\hat{1}-\sum_{\bf k}\hat{G}_{0}({\bf k};i\omega_{n})\hat{U}}.
\end{eqnarray}
%\begin{eqnarray}
%\delta\rho({\bf r},\omega)&=&\delta\rho_{intra}({\bf r},\omega)+\delta\rho_{inter}({\bf r},\omega)\nonumber \\&=&-\frac{1}{\pi}{\rm Im}[\hat{G}_{0}({\bf r};\omega)\hat{T}(\omega)\hat{G}_{0}({\bf r};\omega)]_{11+33}\nonumber \\ &-&\frac{1}{\pi}{\rm Im}[\hat{G}_{0}({\bf r};\omega)\hat{T}(\omega)\hat{G}_{0}({\bf r};\omega)]_{13+31}
%\end{eqnarray}
This yields a correction to the LDOS after the replacement $i\omega_{n}\rightarrow\omega+i0^{+}$,
\begin{eqnarray}
\delta\rho({\bf r},\omega)=-\frac{1}{\pi}{\rm Tr} {\rm Im}\hat{G}_{0}({\bf r};\omega)\hat{T}(\omega)\hat{G}_{0}({\bf r};\omega),
\end{eqnarray}
and its Fourier transform,
%\begin{eqnarray}
%\delta\rho({\bf q},\omega)=\delta\rho_{intra}({\bf q},\omega)+\delta\rho_{inter}({\bf q},\omega)\nonumber \\&=&-\frac{1}{\pi}{\rm Im}[\sum_{\bf k}\hat{G}_{0}({\bf k};\omega)\hat{T}(\omega)\hat{G}_{0}({\bf k+q};\omega)]_{11+33}\nonumber \\ &-&\frac{1}{\pi}{\rm Im}[\sum_{\bf k}\hat{G}_{0}({\bf k};\omega)\hat{T}(\omega)\hat{G}_{0}({\bf k+q};\omega)]_{13+31}
%\end{eqnarray}
\begin{eqnarray}
\delta\rho({\bf q},\omega)=-\frac{1}{\pi}{\rm Tr}{\rm Im}\sum_{\bf k}\hat{G}_{0}({\bf k};\omega)\hat{T}(\omega)\hat{G}_{0}({\bf k+q};\omega).
\end{eqnarray}
These two quantities are proportional to the local differential tunneling conductance as measured in STM, and the QPI signals in phase-sensitive FT-STM, respectively.

We first discuss briefly the LDOS as a function of energy at a single nonmagnetic impurity in real space. Figure 1(d) compares the results for $s^{++}$ and $s^{+-}$ pairing symmetries. For $s^{++}$, the LDOS always shows a typical U-shape feature (inset) within the small gap $\pm|\Delta^{\alpha}|$ regardless of the intraband and interband impurity scattering strength. However, for $s^{+-}$, a similar U-shape feature also appears with Born or weak interband impurity scattering at $v=0.02$ and $u=0.1$ [red solid line in Fig. 1(d)]. Interestingly, for strong interband scattering such as at $v/u=1$ and $u=0.1$ in Fig. 1(d), two intra-gap states arise below the small gap edge $\pm|\Delta^{\alpha}|$ for $s^{+-}$ as a result of its sign-reversing pairing gap between the electron and hole Fermi surfaces. The intensity of the intragap state is much larger at positive energy than at negative energy, and their locations move to lower energies in the unitary limit ($v/u=1$ and $u=10$). The existence of these intragap states, if probed in STM, could provide a smoking gun to distinguish the $s^{++}$- and $s^{+-}$-wave pairing \cite{Hirschfeld2015,Hirschfeld2018,Gao2018}.

We now investigate the QPI for $s^{++}$ and $s^{+-}$ so as to provide characteristic signals for FT-STM measurements. According to the well-known octet model proposed to verify the nodal $d_{x^{2}-y^{2}}$-wave pairing in cuprate superconductors, $\delta\rho({\bf q},\omega)$ should be sensitive to the sign of $\Delta_{\bf k}\Delta_{\bf k+q}$. For certain wave vectors ${\bf q}$, the QPI signal disperses with $\omega$ and features a sharp peak at the resonance energy $\omega<\Delta^{{\rm max}}$ when $\Delta_{\bf k}\Delta_{\bf k+q}<0$. The peak disappears if $\Delta_{\bf k}\Delta_{\bf k+q}>0$. These features are, however, not present for nodeless $s$-wave pairing in iron-based superconductors \cite{Yamakawa2015} and CeCu$_2$Si$_2$. To understand this, we may analyze, for example, the interband scattering contribution to $\delta\rho({\bf q},\omega)$:
\begin{eqnarray}
\delta\rho_{\rm inter}({\bf q},\omega)\propto-{\rm Im}\frac{\omega^{2}-\Delta^{\alpha}\Delta^{\beta}}{(\omega^{2}-(\Delta^{\alpha})^{2})(\omega^{2}-(\Delta^{\beta})^{2})}.
\end{eqnarray}
Hence, for $\omega<|\Delta^{\alpha}|$, the QPI signals always exist because both the numerator and denominator have finite value in spite of the sign of $\Delta^{\alpha}\Delta^{\beta}$; only their peak intensity depends on the sign of $\Delta^{\alpha}\Delta^{\beta}$ and the interband scattering strength.

To see this more clearly, we calculate and plot the QPI spectra $\delta\rho({\bf q},-0.8\Delta^{\alpha})$ at $v/u=1$ and $u=0.1$ in Fig. 2. For intraband scattering as seen in Fig. 2(a), $\delta\rho_{\rm intra}({\bf q},\omega)$ shows no difference between the two pairing symmetries and exhibits bright QPI signals circling around ${\bf q}=(0,0)$. While for interband scattering shown in Fig. 2(b) for $s^{++}$ and Fig. 2(c) for $s^{+-}$, $\delta\rho_{\rm inter}({\bf q},\omega)$ peaks around ${\bf q}=(\pi,\pi)$ and the peak positions barely change with altering energy below $|\Delta^{\alpha}|$, in contrast to those in cuprate superconductors. Because of the opposite gap sign on heavy electron and hole Fermi surfaces ($\Delta^{\alpha}\Delta^{\beta}<0$), the QPI peak intensity of the $s^{+-}$ wave can be much larger than that of the $s^{++}$ wave. However, this is only a quantitative difference. Even with strong interband impurity potential, one could hardly distinguish the $s^{++}$- and $s^{+-}$-wave pairing symmetry solely based on the appearance of the QPI peaks around ${\bf q}=(\pi,\pi)$.

\begin{figure}[t]
\centering\includegraphics[width=0.5\textwidth]{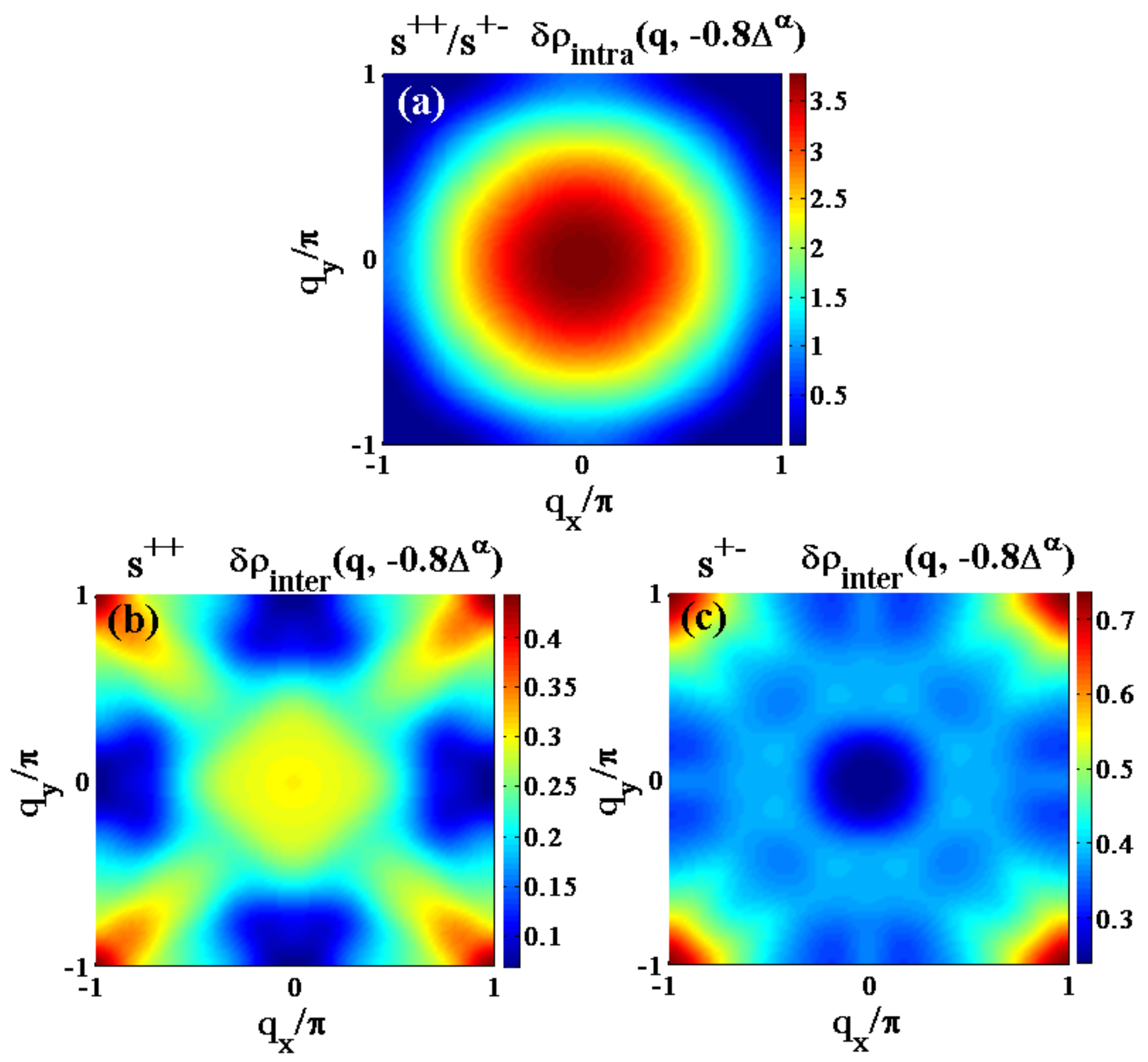}
\caption{Illustration of the QPI spectra at $v/u=1$ and $u=0.1$. (a) $\delta\rho_{\rm intra}({\bf q},-0.8\Delta^{\alpha})$ for both $s^{++}$- and $s^{+-}$-wave pairing. (b) and (c) $\delta\rho_{\rm inter}({\bf q},-0.8\Delta^{\alpha})$ for $s^{++}$-wave and $s^{+-}$-wave pairing, respectively.}
\label{fig2}
\end{figure}
%\textcolor{red}{(Is this revised logic correct?)}

To resolve this issue, we follow the Hirschfeld-Altenfeld-Eremin-Mazin (HAEM) theory and compute the integrated antisymmetric intensity of the correction to the LDOS \cite{Hirschfeld2015},
\begin{eqnarray}
\delta\rho^{-}(\omega)=\delta\rho(\omega)-\delta\rho(-\omega),
\end{eqnarray}
where $\delta\rho(\omega)=\sum_{\bf q}\delta\rho({\bf q},\omega)$. As discussed in the original proposal of Ref. [24], the integrated antisymmetrized LDOS $\delta\rho^{-}(\omega)$ from small ${\bf q}$ centered around $(0,0)$ is contributed mainly by intraband scattering $\delta\rho^{-}_{\rm intra}(\omega)$, which contains no $\Delta^{\alpha}\Delta^{\beta}$ term and is therefore desensitized to the sign change, while that from large ${\bf q}$ around $(\pi,\pi)$ comes mainly from interband scattering \cite{Hirschfeld2015},
\begin{eqnarray}
\delta\rho^{-}_{\rm inter}(\omega)\propto-{\rm Im}\frac{\omega^{2}-\Delta^{\alpha}\Delta^{\beta}}{\sqrt{\omega^{2}-(\Delta^{\alpha})^{2}}\sqrt{\omega^{2}-(\Delta^{\beta})^{2}}}.
\end{eqnarray}
We see immediately some essential distinctions between different pairing states. Although $\delta\rho^{-}_{\rm inter}(\omega)$ are nonzero within the energy interval $[|\Delta^{\alpha}|,|\Delta^{\beta}|]$ for both $s^{+-}$ and $s^{++}$, its intensity is much larger for $s^{+-}$ because $\Delta^{\alpha}\Delta^{\beta}<0$. Most importantly, because $\omega^{2}-\Delta^{\alpha}\Delta^{\beta}$ is always positive, $\delta\rho^{-}_{\rm inter}(\omega)$ retains the same sign between two gaps for $s^{+-}$-wave pairing, while it changes sign for $s^{++}$-wave pairing at $\omega=\sqrt{\Delta^{\alpha}\Delta^{\beta}}$ where $\omega^{2}-\Delta^{\alpha}\Delta^{\beta}$ begins to change the sign. This qualitative difference provides a guide for FT-QPI measurements.

For clarity, we plot $|\delta\rho_{\rm inter}^{-}({\bf q},0.8\Delta^{\alpha})|$ in Fig. 3(a) for $s^{++}$ and Fig. 3(b) for $s^{+-}$  at $v/u=0.2$ and $u=0.1$ with Born or weak interband impurity scattering. While their QPI pattern mainly consists of a weaker QPI signal around ${\bf q}=(\pi,\pi)$ and shows no qualitative difference, a significant distinction can be revealed in Fig. 3(c) in their integrated antisymmetrized LDOS $\delta\rho^{-}_{\rm inter}(\omega)$ as a function of energy. It is clear that the sign of  $\delta\rho^{-}_{\rm inter}(\omega)$ stays the same within the energy window between $|\Delta^{\alpha}|$ and $|\Delta^{\beta}|$ for $s^{+-}$ but changes for $s^{++}$. One may thus expect that analyses of the integrated antisymmetrized LDOS in phase-sensitive QPI measurements can examine the sign structure of the pairing symmetry in CeCu$_2$Si$_2$. As a matter of fact, a similar technique has proved quite successful in LiOH-intercalated FeSe and NaFe$_{1-x}$Co$_x$ \cite{Du2017,Cheung2020}.

\begin{figure}[t]
\centering\includegraphics[width=0.5\textwidth]{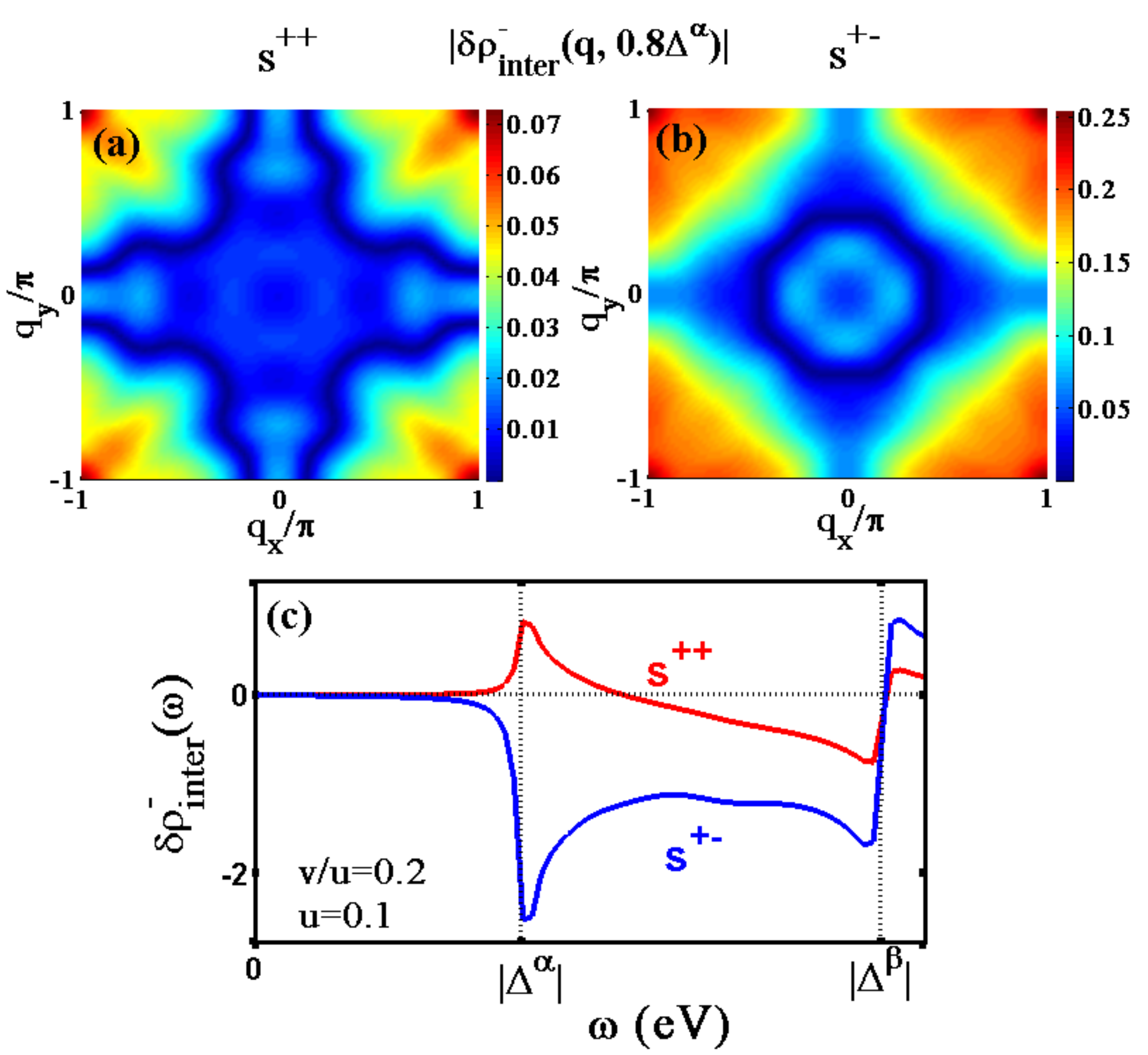}
\caption{Illustration of the QPI spectra for weak interband impurity scattering at $v/u=0.2$ and $u=0.1$. (a) $|\delta\rho_{\rm inter}^{-}({\bf q},0.8\Delta^{\alpha})|$ for the $s^{++}$ wave; (b) $|\delta\rho_{\rm inter}^{-}({\bf q},0.8\Delta^{\alpha})|$ for the $s^{+-}$ wave; (c) comparison of the integrated antisymmetrized LDOS $\delta\rho^{-}_{\rm inter}(\omega)$ for $s^{++}$ and $s^{+-}$. }
\label{fig3}
\end{figure}

The situation is similar for strong interband impurity scattering. Figure 4 compares the total integrated antisymmetrized LDOS $\delta\rho^{-}(\omega)$ at $v/u=1$ and $u=0.5$ with both intraband and interband contributions. The interband component $\delta\rho^{-}_{\rm inter}(\omega)$ is also plotted in the inset and is seen to behave similarly and play a major role in the total $\delta\rho^{-}(\omega)$. Two essential features in $\delta\rho^{-}(\omega)$ can be used to distinguish the $s^{++}$- and $s^{+-}$-wave pairing states. First, for $s^{+-}$, a sharp impurity resonance peak appears at $\omega<|\Delta^{\alpha}|$ due to strong interband impurity scattering, which corresponds to the intragap impurity resonance states as shown in Fig. 1(d). Second, $\delta\rho^{-}(\omega)$ within the energy window $[|\Delta^{\alpha}|,|\Delta^{\beta}|]$ exhibits the same property as in Fig. 3(c) for weak interband impurity scattering. That is to say, $\delta\rho^{-}(\omega)$ changes sign for $s^{++}$ but has no sign change for $s^{+-}$. This feature is always preserved even if we eliminate the effect of the impurity resonance peak for $s^{+-}$ as shown in the filtered $\delta\rho^{-}(\omega)$ in Fig. 4. Therefore the integrated antisymmetrized LDOS provides a robust criterion for revealing the sign reversal of the pairing symmetry, no matter whether there is an intra-gap impurity resonance or not.

Altogether, we may give some definitive criteria for distinguishing the sign-reversing $s^{+-}$-wave pairing from the sign-preserving $s^{++}$-wave pairing to guide the STM and phase-sensitive QPI measurements in CeCu$_2$Si$_2$: (1) The appearance of intra-gap states in LDOS and antisymmetrized LDOS $\delta\rho^{-}(\omega)$, if observed within the experimental energy $\omega<|\Delta^{\alpha}|$, is a smoking gun for $s^{+-}$-wave pairing. (2) The sign of the integrated antisymmetrized LDOS $\delta\rho^{-}_{\rm inter}(\omega)$ and $\delta\rho^{-}(\omega)$, if unchanged between the small gap $|\Delta^{\alpha}|$ and the large gap $|\Delta^{\beta}|$, can unambiguously verify the $s^{+-}$-wave pairing, regardless of the strength of interband impurity scattering. (3) Only the QPI signal $\delta\rho({\bf q},\omega)$ is not sufficient to distinguish $s^{+-}$ and $s^{++}$ for CeCu$_2$Si$_2$ with nodeless $s$-wave pairing.

\begin{figure}[t]
\centering\includegraphics[width=0.45\textwidth]{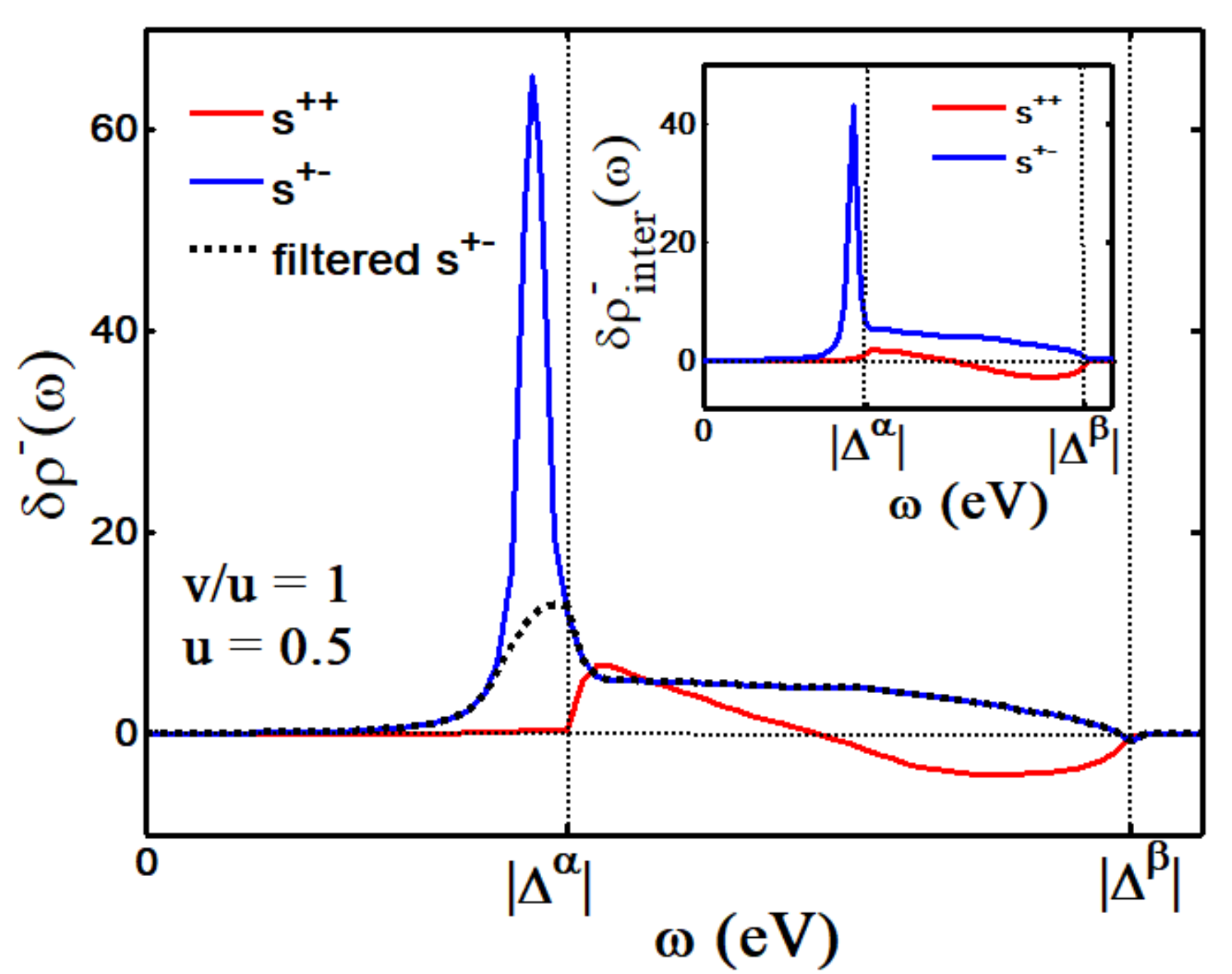}
\caption{Comparison of the total integrated antisymmetrized LDOS $\delta\rho^{-}(\omega)$ for $s^{++}$ and $s^{+-}$ with strong interband impurity scattering at $v/u=1$ and $u=0.5$. The dotted line is the filtered $\delta\rho^{-}(\omega)$ for $s^{+-}$ after removing the effect of the impurity resonance peak. The inset compares their interband scattering contributions $\delta\rho^{-}_{\rm inter}(\omega)$.}
\label{fig4}
\end{figure}

In conclusion, we have investigated the LDOS and QPI signals induced by weak and strong interband impurity scattering by applying the standard $T$-matrix approach with realistic multi-band structures from the density functional theory with Hubbard $U$ correction (DFT+$U$) for CeCu$_2$Si$_2$. We find decisive distinctions between $s^{+-}$ and $s^{++}$ pairing states, which include the intra-gap impurity resonance state below the small gap energy and a significant energy-dependent difference in the sign of the integrated antisymmetrized LDOS between the small and large gap energies. Our work provides a useful guide for future STM and phase-sensitive QPI measurements to identify the true pairing symmetry in the heavy-fermion superconductor CeCu$_2$Si$_2$.

This work was supported by the National Natural Science Foundation of China under Grants No. 11774025, No. 11774401, No. 11974051, and No. 11734002, the National Key R\&D Program of MOST of China (Grant No. 2017YFA0303103), and the Strategic Priority Research Program of the Chinese Academy of Sciences (Grant No. XDB33010100).

\end{document}